\documentclass[twocolumn,showpacs,aps,preprintnumbers,floatfix,amsmath,amssymb,prc]{revtex4-1}
\usepackage{bm}% bold math
\usepackage{graphicx}% Include figure files
\usepackage{subfigure}

\newcommand{\ve}[1]{\ensuremath{\mathbf{#1}}}
\newcommand{\n}[1]{\ensuremath{|\mathbf{#1}|}}
\newcommand{\scp}[2]{\ensuremath{\mathbf{#1}\cdot\mathbf{#2}}}
\newcommand{\vep}[2]{\ensuremath{\mathbf{#1}\times\mathbf{#2}}}
\newcommand{\Ek}{\ensuremath{E_k}}
\newcommand{\Ep}{\ensuremath{E_p}}

\newcommand{\Epp}{\ensuremath{E_{p'}}}
\def\lsim{\lesssim}
\def\gsim{\gtrsim}
\def\beq{\begin{equation}}
\def\eeq{\end{equation}}
\def\be{\begin{eqnarray}}
\def\ee{\end{eqnarray}}

\begin{document}

\title{Electroweak nuclear response at moderate momentum transfer}

\author{Artur M. Ankowski}
\altaffiliation[On leave from ]{Institute of Theoretical Physics, University of Wroc{\l}aw, Wroc{\l}aw, Poland}
\email{Artur.Ankowski@roma1.infn.it}
\affiliation{INFN and Department of Physics,``Sapienza'' Universit\`a di Roma, I-00185 Roma, Italy}
\author{Omar Benhar}
\affiliation{INFN and Department of Physics,``Sapienza'' Universit\`a di Roma, I-00185 Roma, Italy}

\date{\today}

\graphicspath{{plot/}}

\begin{abstract}
We discuss the convergence of the expansion of the nuclear electroweak current in powers of
$|{\bf k}|/M$, where $M$ is the nucleon mass and ${\bf k}$ denotes either the momentum transfer
or the momentum of the struck nucleon.
We have computed the electron and neutrino scattering cross sections off uniform nuclear matter at
equilibrium density using correlated wave functions and the cluster expansion formalism.
The results of our work suggest that the proposed approach provides accurate estimates of the
measured electron scattering cross sections. On the other hand, the description of the current based on the
widely used leading-order approximation does not appear to be adequate, even at momentum transfer as low
as 300 MeV.
\end{abstract}

%Nuclear matter
%\keywords{Suggested keywords}%Use showkeys class option if keyword display desired
\maketitle

\section{Introduction}
Many aspects of the recent neutrino data remain puzzling~\cite{ref:MiniB_kappa, ref:MINOS_QE,ref:MiniB_excess,ref:MINOS_asymm}. While unconventional explanations have been proposed~\cite{ref:Akhmedov,ref:Kopp,ref:Gninenko}, before advocating exotics, more conventional
effects should be analyzed within realistic models.

One of the main sources of uncertainty in the interpretation of the available data is the treatment of nuclear effects,
which are often described using oversimplified models of both nuclear dynamics and the reaction mechanism.

Most neutrino experiments take data at beam energy $E_\nu \lsim 1$ GeV, where quasielastic scattering
dominates. In this channel the elementary neutrino-nucleon cross section is determined by the form factors
$F^1_{N}$, $F^2_{N}$, and $F_A$, where $N=p$ and $n$ for protons and neutrons, respectively. The vector
form factors have been measured to high accuracy in electron-proton and electron-deuteron scattering experiments, whereas
the $Q^2$ dependence of the axial form factor $F_A$ is still controversial, as different collaborations reported
significantly different values of the axial mass $M_A$~\cite{ref:MiniB_CS, ref:NOMAD}.

Many studies of neutrino-nucleus scattering are based on the impulse approximation (IA), the underlying
assumptions of which are that (i) the momentum transfer is large enough for the probe to interact with only one nucleon, and (ii) final-state
interactions, of both statistical and dynamical origin,  between the struck particle and the spectators can be neglected.

In a previous paper~\cite{ref:ABF10}, we investigated the limits of applicability
of the IA scheme in the case of the linear response of infinite nuclear
matter to a scalar probe. The results of that analysis suggest that, at momentum transfer below about twice the Fermi momentum, corresponding to
$\sim$500 MeV at equilibrium density, the IA breaks down, mostly due to the oversimplified description of the final state.

In this paper, we develop a formalism suitable to obtain realistic estimates of the lepton-nucleus cross section in
the region of moderate momentum transfer. As an illustrative example, we consider uniform nuclear matter,
consisting of equal numbers of protons and neutrons, at equilibrium density ($\rho = 0.16 \ {\rm fm}^{-3}$,
corresponding to Fermi momentum $k_F = 1.33 \ {\rm fm}^{-1}$).

\section{Electroweak cross section}
The scattering cross
section can be written as
\begin{equation}\label{eq:xsec}
d \sigma \propto L_{\mu \nu} W^{\mu \nu},
\end{equation}
where $L_{\mu \nu}$ is determined by the lepton kinematics and
\begin{equation}\label{eq:response}
W^{\mu \nu} = \sum_n \langle 0 | {J^\mu}^\dagger | n \rangle \langle n | J^\nu | 0 \rangle
\: \delta(\omega + E_0 - E_n).
\end{equation}
In the above equation, $\omega$ is the energy transfer, $| 0 \rangle$ and $| n \rangle$ are the initial and final
states of the target, and $J_\mu$ is the current operator accounting for the lepton-nucleus interaction.
Within our approach, nuclear matter is described using {\em correlated}
states, including the effects of the nonperturbative components of nucleon-nucleon (NN) forces \cite{ref:Feenberg,ref:Clark}.

We have used Eqs.~\eqref{eq:xsec} and~\eqref{eq:response} to obtain the electron and neutrino
scattering cross sections, including the contributions arising from one-particle--one-hole correlated states. In our calculation,
antisymmetrization of the final state and dynamical final-state interactions are both taken into account.

At moderate momentum transfer $\n q \lsim 500$ MeV, the nuclear current operator
\begin{equation}
J_\mu = \sum_i j^i_\mu,
\end{equation}
where $j_\mu$ is the nucleon current, is usually treated expanding in powers of the ratio
$|{\bf k}|/M$ and keeping only the leading terms. Here, $M$ is the nucleon mass, whereas ${\bf k}$ denotes
either the momentum transfer or the momentum of the struck nucleon.

%However, the corrections tothe commonly used lowest order approximation turn out to be large.
In this paper, we present the results of a systematic analysis
%Hence, we carried out a systematic analysis of
of the convergence of this expansion, showing that contributions beyond leading order may be important and cannot be neglected.
A similar analysis, carried out within the context
of the mean-field approach using the formalism reviewed in Ref.~\cite{Amaro02}, is discussed in Ref.~\cite{Amaro96}.

\section{Nuclear Hamiltonian and nuclear wave functions}
The correlated states are defined as \cite{ref:Feenberg,ref:Clark}
\begin{equation}\label{eq:corrStates}
| n \rangle = \frac{ F | n ) }
{ ( n | F^\dagger F | n )^{1/2} },
\end{equation}
where $| n )$ is a Fermi gas (FG) state and the operator $F$, embodying the correlation structure
induced by the NN interaction, is written in the form
\begin{equation}
F(1,\ldots,N)=\mathcal{S}\prod_{j>i=1}^N f_{ij},
\label{def:F}
\end{equation}
where $\mathcal{S}$ is the symmetrization operator, accounting for the fact that, in general,
$\left[ f_{ij} , f_{ik} \right] \neq 0$.

The two-body correlation function $f_{ij}$ features an operatorial structure reflecting
the complexity of the NN potential:
\begin{equation}
f_{ij} = \sum_{TS} \left[ f_{TS}(r_{ij}) + \delta_{1,\,S}\:f_{tT}(r_{ij}) S_{ij} \right] \: P_{TS}.
\end{equation}
In the above equation, $r_{ij} = |{\bf r}_{ij}|$ with ${\bf r}_{ij}={\ve r}_i - {\ve r}_j$, $P_{TS}$ is the operator projecting onto
two-nucleon states of total spin and isospin $S$ and $T$, respectively, and
\begin{equation}
S_{ij}=3({\bm \sigma}_i\cdot{\bf r}_{ij})
({\bm \sigma}_j \cdot {\bf r}_{ij})/r_{ij}^2-({\bm \sigma}_i \cdot {\bm \sigma}_j).
\end{equation}
The shapes of the radial functions $f_{TS}(r_{ij})$ and $f_{tT}(r_{ij})$, are determined
through functional minimization of the expectation value of the nuclear Hamiltonian in
the correlated ground state. The assumption that the correlation operator obtained
using this procedure can be used to generate the whole basis of correlated states is one of the main tenets
 of correlated-basis function (CBF) perturbation theory.  It relies on the
premise that the mean field and the correlation structure induced
by the nucleon-nucleon interaction are largely decoupled from
one another. The validity of this assumptions is supported
by the fact that the results of CBF calculation of a variety of
nuclear matter properties are in quantitative agreement with
the results of Monte Carlo calculations, as well as
with the available empirical data.

The calculations of matrix elements of both the Hamiltonian and the nuclear current
between correlated states involve severe difficulties. To overcome this problem, in our work, we use the cluster expansion
formalism~\cite{ref:Clark}, which amounts to writing the matrix element as a sum of contributions arising from
subsystems involving an increasing number of nucleons.

The results discussed in the present paper have been obtained at lowest order, i.e. including two-body cluster
 contributions only.
Using this approximation appears to be justified in the context of a calculation
of the nuclear matter response at equilibrium density~\cite{ref:CP04,ref:BF09}.

The Hamiltonian employed to determine the correlation functions includes the kinetic energy term and
the Argonne $v_6$ NN potential~\cite{ref:Argonne_v6}. The resulting correlation functions are then used to construct the
effective interaction discussed in Ref.~\cite{ref:BV07} needed to obtain the nuclear matter
single-particle spectrum~\cite{ref:BF09}. Note that the effective interaction of Ref.~\cite{ref:BV07} includes the effects
of three-nucleon forces, which are known to be important.

\section{Current operator}
The matrix representation of the weak current is the sum of the vector part,
\begin{equation}
\varGamma^\mu_V=\gamma^\mu(F^1+F^2)-\frac{(p+p')^\mu F^2}{2M},
\end{equation}
and the axial one,
\begin{equation}
\varGamma^\mu_A=\gamma^\mu\gamma_5 F_A+\gamma_5\frac{q^\mu F_P}M.
\end{equation}
In general, the upper $\chi_\sigma$ and lower Pauli spinor $\phi_\sigma$ describing an on-shell Dirac particle of mass $M$ and momentum $\ve p$ are related through
\begin{equation}
\phi_\sigma=\ve{\hat p}\cdot\sigma\chi_\sigma,
\end{equation}
where $\ve{\hat p}$ is a compact notation for $\ve p/(M+\Ep)$, with $\Ep=\sqrt{\ve p^2+M^2}$. This relation allows one to reduce the current according to
\begin{equation}\begin{split}
&\Big(\chi_{\sigma'}^\dagger,\; \phi_{\sigma'}^\dagger\Big)\big(\varGamma^\mu_V+\varGamma^\mu_A\big)\begin{pmatrix}\chi_{\sigma}\\ \phi_{\sigma}\end{pmatrix}\rightarrow\chi_{\sigma'}^\dagger\big(V^\mu+A^\mu\big)\chi_\sigma,
\end{split}\end{equation}
with
\begin{equation}\begin{split}
\frac{V^0}{\lambda_p\lambda_{p'}}&=\big(F^1+F^2\big)\big[1+\scp{\hat p'}{\hat p}+i\sigma\cdot(\vep{\hat p'}{\hat p})\big]\\
&\quad-F^2\frac{\Epp+\Ep}{2M}\big[1-\scp{\hat p'}{\hat p}-i\sigma\cdot(\vep{\hat p'}{\hat p})\big],\\
\frac{V^k}{\lambda_p\lambda_{p'}}&=\big(F^1+F^2\big)\big[(\ve{\hat p'}+i\sigma\times\ve{\hat p'})^k+(\ve{\hat p}-i\sigma\times\ve{\hat p})^k\big]\\
&\quad-F^2\frac{(\ve{p}+\ve{p'})^k}{2M}\big[1-\scp{\hat p'}{\hat p}-i\sigma\cdot(\vep{\hat p'}{\hat p})\big],\\%
\frac{A^0}{\lambda_p\lambda_{p'}}&=F_A\:\sigma\cdot\big(\ve{\hat p'}+\ve{\hat p}\big)-F_P\frac{\Epp-\Ep}{M}\:\sigma\cdot\big(\ve{\hat p'}-\ve{\hat p}\big),\\
\frac{A^k}{\lambda_p\lambda_{p'}}&=F_A\big[\sigma^k+\hat p'^k(\sigma\cdot\ve{\hat p})+\hat p^k(\sigma\cdot\ve{\hat p'})-(\scp{\hat p'}{\hat p})\:\sigma^k\big]\\
&\quad-iF_A(\vep{\hat p'}{\hat p})^k-F_P\frac{(\ve{p'}-\ve{p})^k}{M}\:\sigma\cdot\big(\ve{\hat p'}-\ve{\hat p}\big),\\
\end{split}\end{equation}
where $\lambda_k=\sqrt{(\Ek+M)/2\Ek}$ and $k=p,\ p'$.

The standard approach is to expand the above currents in powers of momentum over mass, keeping only the leading order terms:
\begin{equation}\begin{split}
{V^0}&=F^1,\\%
{V^k}&=\big(F^1+F^2\big)\frac{i\big[\sigma\times(\ve{p'}-\ve{p})\big]^k}{2M}+F^1\frac{(\ve{p}+\ve{p'})^k}{2M},\\%
{A^0}&=F_A\frac{\sigma\cdot(\ve{p'}+\ve{p})}{2M}-F_P\frac{\ve{p'}^2-\ve{p}^2}{2M^2}\:\frac{\sigma\cdot\big(\ve{p'}-\ve{p}\big)}{2M},\\
{A^k}&=F_A\sigma^k-F_P\frac{(\ve{p'}-\ve{p})^k}{M}\:\frac{\sigma\cdot\big(\ve{p'}-\ve{p}\big)}{2M}.\\
\end{split}\end{equation}
Within the considered kinematical region, the terms containing $F_P$ and $F_A$ in
the above equations are of the same order, due to the hypothesis of the partially conserved axial current relating the pseudoscalar form factor to the axial
one through  $F_P=2M^2F_A/(m_\pi^2+Q^2)$, where $m_\pi$ is the pion mass.

Note that, in our treatment, off-shell effects are not taken into account in the calculation of the current matrix elements in Eq.~\eqref{eq:response}, but
the single-particle spectrum, needed to obtain the energy of the one-particle--one-hole states, is modified to take into account interactions effects.
As a consequence, we can consistently use free-space nucleon form factors and the kinematics
involves no ambiguities. A~similar, although oversimplified, procedure is used in
calculations based on the Fermi gas model, in which binding is
described through an average nucleon separation energy, entering
the argument of the energy conserving $\delta$ function only.

In the following, we compare the results obtained at leading (LO) and next-to-leading (NLO) order to the exact result for momentum transfer up to $\sim$600 MeV.

%%%%%%%%%%%%%%%%
\begin{figure}
    \centering
    \subfigure{\label{fig:2020@15NM}}
    \includegraphics[width=0.80\columnwidth]{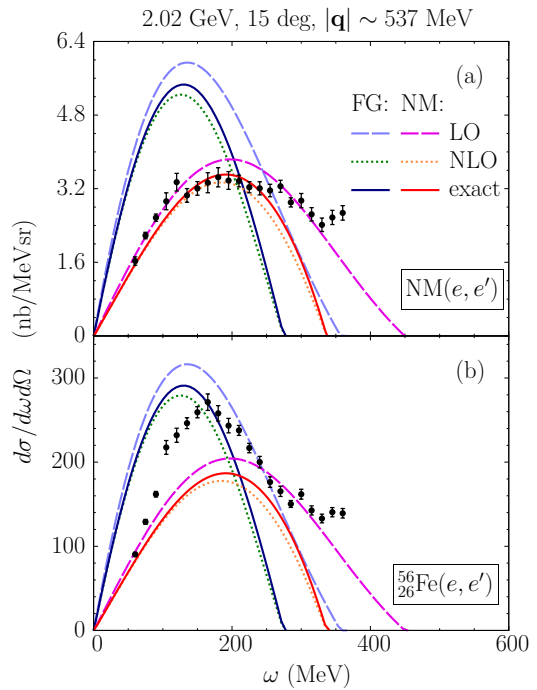}
    \subfigure{\label{fig:2020@15Fe}}
\caption{ (Color online) Panel (a): Differential cross section ${d\sigma}/{d\omega d\Omega}$ for electron scattering off nuclear matter at beam energy 2.02~GeV and scattering angle 15 degrees. The experimental data are taken form Ref.~\cite{ref:Day&al_NM}.
%are not corrected for nucleon excitations.
Panel (b): Comparison of the nuclear matter cross sections with the $^{56}_{26}$Fe$(e,e')$ data form Ref.~\cite{ref:Day&al_Fe}, collected at the same kinematics. Calculations include the correction for neutron excess in the iron nucleus.}
\end{figure}
%%%%%%%%%%%%%%%%

%%%%%%%%%%%%%%%%
\begin{figure*}
%%%%%%%%%%%%%%%%
    \centering
    \includegraphics[width=0.90\textwidth]{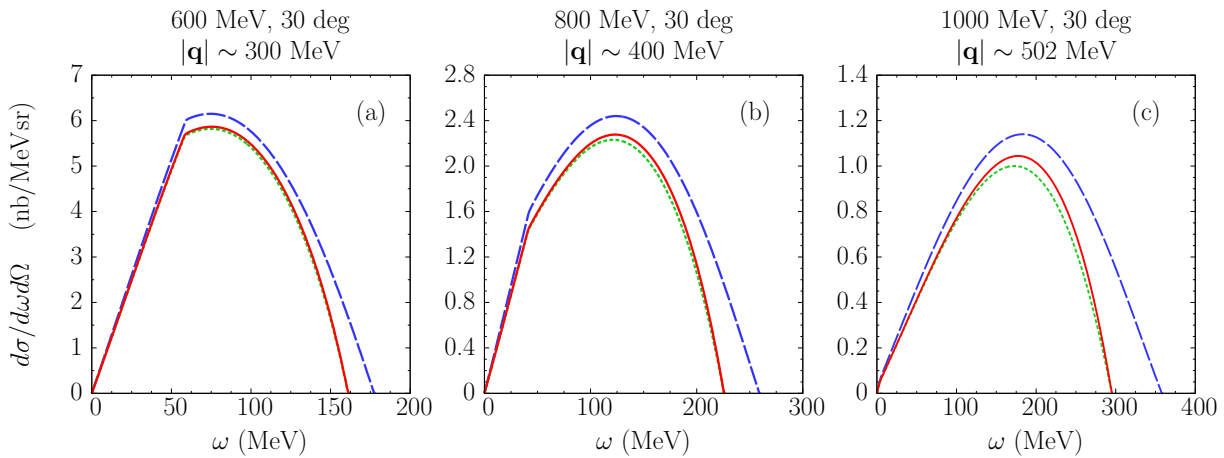}
%%%%%%%%%%%%%%%%
\caption{\label{fig:electrons} (Color online) Differential cross section ${d\sigma}/{d\omega d\Omega}$ for electron scattering off nuclear matter calculated exactly (solid line), in leading order (dashed line), and in next-to-leading order (dotted line). The panels are labeled with beam energy and scattering angle. The values of the momentum transfer refer to the quasielastic peak.}
\end{figure*}
%%%%%%%%%%%%%%%%

%%%%%%%%%%%%%%%%
\begin{figure*}
%%%%%%%%%%%%%%%%
    \centering
    \includegraphics[width=0.90\textwidth]{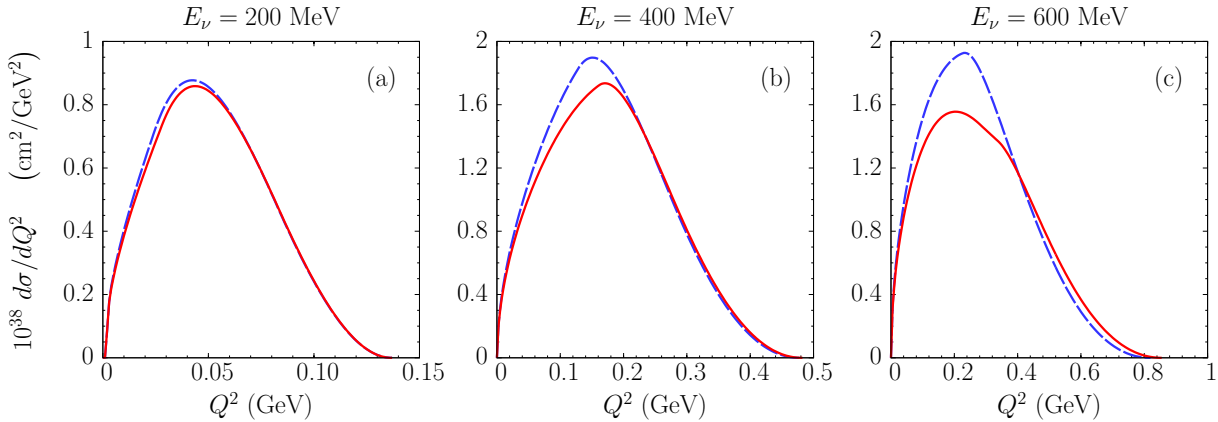}
%%%%%%%%%%%%%%%%
\caption{\label{fig:neutrinos} (Color online) Comparison of the differential cross section ${d\sigma}/{dQ^2}$ for muon neutrino scattering off nuclear matter calculated exactly (solid line) and in leading order (dashed line).}
\end{figure*}
%%%%%%%%%%%%%%%%

%The standard approach is to expand the above currents in powers of momentum over mass, keeping only the leading order terms. In the following, we compare
%the results obtained at leading (LO) and next-to-leading (NLO) order to the exact result for momentum transfer up to $\sim$600 MeV.

\section{Results}
Figure~\ref{fig:2020@15NM} shows a comparison between our results and the inclusive electron scattering cross section of nuclear matter at $\n q=537$ MeV, obtained in Ref.~\cite{ref:Day&al_NM} by extrapolating the available data for finite nuclei. The curve labeled LO corresponds to nonrelativistic kinematics and nonrelativistic single-particle
spectrum, while the one labeled NLO has been obtained using relativistic kinematics and kinetic energies, and including next-to-leading-order terms in the expansion of the current.
The exact calculation includes all terms. Note that most existing studies of the nuclear matter response carried out within nuclear many-body theory using
CBFs are based on the LO approximation~\cite{ref:Fabrocini&Fantoni,ref:Fabrocini}. The results of Fig.~\ref{fig:2020@15NM} show that, even at momentum transfer as low as
$\sim$500 MeV,  the contribution of higher-order terms is not negligible, and must be taken into account.
It clearly appears that the NLO and exact calculations accurately reproduce the data in the region of the quasielastic peak.
For comparison, we also show the corresponding results obtained using the FG model with pure kinetic energy spectrum.
Note that, while the wrong position of the maximum of the FG cross section might be fixed introducing an average nucleon separation
energy, the failure to reproduce height and width of the measured cross section cannot be cured without an {\em ad hoc} modification
of the Fermi momentum.

To illustrate the importance of finite size effects, in Fig.~\ref{fig:2020@15Fe} we compare our nuclear matter results, in the same kinematical setup as in Fig.~\ref{fig:2020@15NM}, to the cross section measured using an iron target~\cite{ref:Day&al_Fe}. The theoretical results have been corrected to take into account neutron excess in iron, the influence of which turns out to be negligibly small. It clearly appears that finite size effects are large and that a meaningful comparison can only be done using extrapolated data.
Unfortunately, however, the extrapolation procedure developed in Ref.~\cite{ref:Day&al_NM} is based on the IA scheme, and cannot be employed for momentum transfer below $\sim$500 MeV~\cite{foot1}.

In the following, we only discuss the results of nuclear matter calculations, for both electron and neutrino scattering, to
illustrate the convergence of the expansion of the current operator in different kinematical conditions.

Figure~\ref{fig:electrons} shows the electron scattering cross sections at fixed scattering angle, 30 degrees, and beam energies ranging between
600 and 1000 MeV, corresponding to momentum transfer $300 \lsim \n q \lsim 500$ MeV.  Comparison between the dashed line, corresponding
to the LO calculation, and the solid line, representing the exact result, shows that even at momentum transfer as low as 300 MeV, the effect of using
relativistic kinematics  is sizable. Contrary to the nonrelativistic case, when relativistic kinematics is used, the width of the nuclear matter response becomes independent of $\n q$ at large momentum transfer, being determined by the Fermi momentum only. A suppression of the cross section is also clearly visible
over the whole energy loss range.

To make a connection with an observable that is often analyzed in neutrino experiments, in Fig.~\ref{fig:neutrinos} we show the $Q^2$ distribution at fixed beam energies $200 \leq E_\nu \leq 600$~MeV. The LO approximation appears to be quite accurate at the lowest neutrino energy, where the cross section
picks up contributions at momentum transfer $170 \lsim \n q \lsim 300$ MeV. As $E_\nu$
increases, inclusion of higher-order terms leads to the appearance of sizable corrections, affecting both the magnitude and the shape of the
$Q^2$ distributions.

\section{Conclusions}
We have developed a theoretical approach allowing for a consistent treatment of electron- and neutrino-nucleus interactions in the region of moderate momentum transfer, in which the IA is expected to break down. Our results suggest that the widely used LO approximation
for the nonrelativistic reduction of the nuclear current conspicuously fails at momenta $\gsim$ 300 MeV. On the other hand, inclusion of higher-order
terms leads to a remarkably good agreement between theory and extrapolated electron scattering data at $\n q\sim500$~MeV.

The main feature of our work is the inclusion of nu{\-}cle{\-}on-nucleon correlations, the effects of which are long known to be important in electron-nucleus
scattering~\cite{RMP1}. The use of nonrelativistic correlation functions in the final state may be questionable at large momentum transfer.
However, as $\n q$ increases the role of correlations between the struck nucleon, carrying a large momentum, and the spectators becomes less and less
important, and the impulse approximation limits is smoothly recovered~\cite{ref:ABF10}.

A systematic comparison with the available data will require the extension of our formalism to the case of finite nuclei. In addition, the contribution of
processes involving meson-exchange currents, which is known to be significant in the transverse electromagnetic response~\cite{ref:Carlson&Schiavilla98}, needs
to be included.

The approach based on correlated basis states, discussed in this  paper, can be readily generalized to describe finite systems. It has been already applied
to obtain a variety of properties of medium-heavy nuclei ($^{12}$C, $^{16}$O, $^{40}$Ca, and $^{48}$Ca), such as ground-state energies, charge-density and momentum distributions, natural orbits, occupation numbers, quasihole wave functions, and spectroscopic factors~\cite{ABCF}. The extension of the techniques developed in
Ref.~\cite{ABCF} to the calculation of transition matrix elements of the electroweak current does not involve any additional conceptual problems.

The role of meson-exchange currents in neutrino-nucleus interactions has long been recognized \cite{singh}. Their effects must be included in a consistent fashion, as the
longitudinal components are related to the nuclear Hamiltonian, determining the correlation
functions, through the continuity equation. The formalism to obtain the model-independent part of the currents within nonrelativistic many-body theory is discussed
in Ref.~\cite{SPR}, while the relativistic expressions, as well as their reduction, are analyzed in Ref.~\cite{Amaro02}. It has been recently suggested
\cite{ref:ABF10,martini,newpar,barbaro} that the effects of
meson-exchange  currents may explain at least part of the discrepancies between the results of theoretical calculations and the double-differential charged-current
quasielastic cross section measured by the MiniBooNE Collaboration~\cite{ref:MiniB_CS}.

%Therefore, comparison between nuclear matter calculations and Iron data discussed in the following is mainly meant to provide a qualitative picture of the %main features of the cross section and analyze the importance of relativistic kinematics and higher order terms in the expansion of the nuclear current at %lower momentum transfer.

%\section{Cluster expansion of the nuclear electroweak response}

%\section{Results}

%\section{Summary and Outlook}

%\begin{itemize}

%\item for neutrino scattering, the exact calculation is much easier than the NLO

%\item briefly discuss the issue of gauge invariance in connection with
%two-body current contribution.

%\end{itemize}

\begin{acknowledgments}
A.M.A. was supported by the Polish Ministry of Science and Higher Education under Grant No. 550/MOB/ 2009/0. This work was supported by INFN under Grant MB31.
\end{acknowledgments}


\begin{thebibliography}{99}
\bibitem{foot1}%
    {Ref. \cite{ref:Day&al_NM} also provides the extrapolated cross section at beam energy 500 MeV and scattering angle 60 degrees,
    corresponding  to momentum transfer $\sim$447 MeV. However, due to the large scattering angle, this cross section is mostly transverse, and dominated
    by $\Delta$ production.}
\end{thebibliography}
\end{document}